# The modeling of the ENSO events with the help of a simple model

Vladimir N. Stepanov

Proudman Oceanographic Laboratory, Merseyside, England

February 20, 2007

## Abstract

The El Niño Southern Oscillation (ENSO) is modelled with the help of a simple model representing a classical damped oscillator forced by external forcing. Eastern Pacific sea surface temperature (SST) and the mean equatorial Pacific thermocline depth correspond to the roles of momentum and position. The external forcing of the system is supplied by short-period meridional mass fluctuations in the Pacific sector of the Southern Ocean due to the joint effect of the atmospheric variability over the Antarctic Circumpolar Current (ACC), bottom topography and coastlines, and also by the variability of westerly winds in the tropics. Under such conditions the ENSO-like oscillations arise as a result of propagation of signals due to both initial signals appeared in the Southern Ocean and the tropical westerly wind anomaly, that propagate then across the equatorial Pacific by means of fast wave processes. The external forcings are the main factor in establishing the oscillation pattern.

## 1. Introduction

Alvarez-Garcia et al. 2006 from a coupled global climate model simulation have identified three classes of ENSO events. The first two classes are characterized by well known paradigms: the first is the delayed oscillator where equatorial coupled waves produce a delayed-negative feedback to the warm sea surface temperature SST anomalies in the tropics (e.g., see Suarez and Schopf 1988); the second model is the recharge oscillator where fast wave processes adjust the thermocline tilt as a result of wind stress variability (see, e.g., Jin 1996). However, Kessler 2002, argued that ENSO events are like disturbances with respect to a stable basic state, requiring an initiating impulse not contained in the dynamics of the cycle itself, and the initiation might be carried out by some other climate variation. The third class of ENSO events identified by Alvarez-Garcia et al. 2006 is characterized by a relatively quick development of ENSO events (less than 9 months after the changes of the above mentioned plausible exciting forcings in the tropics) and it supports the conclusion of Kessler 2002 about some initiating impulse. This article considers the variability arising from the joint effect of bottom topography, coastlines and atmospheric conditions over the ACC as an amplifier/trigger for ENSO events.

Numerical experiments with the help of a barotropic ocean model (Stepanov and Hughes 2004) forced by 6-hourly global atmospheric winds and pressures from the European Centre for Medium-Range Weather Forecasts (ECMWF), have demonstrated that changes in wind strength over the ACC together with the effect of bottom topography induce some pressure/density anomalies in the Southern Ocean (Stepanov and Hughes 2006; Stepanov, submitted manuscript 2006 (see, also http://arxiv.org/abs/physics/0702159 , henceforth S06). These anomalies lead to short-period variations of meridional flows in the Pacific sector of the Southern Ocean to the north of $47^oS$, that result in the mean value of daily water mass variability $M(t)$ in the Pacific about of 500 Gt (Gigatonns) for a summer period preceding the cold or warm ENSO events (although the real water exchange between the Pacific and the Southern Ocean is more than an order of magnitude higher because of the link between the total meridional flux and mass flux in the western portion of the Pacific obtained in these numerical experiments). These mass variations are anticorrelated with the strength of the wind over the ACC at the 99% confidence level. All these above described features are clear seen from Figure 1, where the transport through Drake Passage and the daily mass variability in the Pacific Ocean $M(t) \sim \int_o^t Q_P(t)\, dt$ due to meridional transport fluctuations $Q_P$ through the latitude of 40°S, averaged for July-September from the model's 20-year time series, are shown. There is a strong coincidence between the minimums and maximums of this mass variability in the Pacific, and cold and warm ENSO events, respectively, i.e., short-term fluctuations in *M(t)* are related to the onset of ENSO events (the correlation coefficient between the mean summer's mass variability *M(t)* presented on Fig. 1 and the winter's NINO4-index (dashed line on Fig. 1) is 0.84 at the 99% confidence level).

This meridional flux variability in the Pacific sector of the Southern Ocean can induce some short-period density anomalies in the vicinity of these regions with high meridional flux variability at the time scales of several months when the effect of barotropic changes has not yet transformed into baroclinic ones. These anomalies can be transferred to low latitudes by the wave mechanism described by Ivchenko et al. 2004, henceforth IZD04, that subsequently interact with stratification changing the tropical thermocline tilt which can amplify an ENSO event. IZD04 showed that signals due to salinity anomalies generated near to Antarctica can propagate almost without changes of disturbance amplitude in the form of fast-moving barotropic Rossby waves. Such waves propagate from Drake Passage (where the ACC is constricted to its narrowest meridional extent) to the western Pacific and are reflected at the western boundary of the Pacific before moving equatorwards and further northwards along the coastline as coastally trapped Kelvin waves. Such signals propagate from Drake Passage to the equator in only a few weeks and through the equatorial region in a few months (henceforth denote via $T_B$ the period needed for anomalies from the Southern Ocean to reach low latitudes). In a more realistic coupled ocean-atmosphere general circulation model, Richardson et al. 2005 observed a similar rapid response of the Pacific to a similar density anomaly in the ACC.

The above described mass variability in the Southern Ocean can significantly influence the tropics. The value of daily mass variability in the Pacific about 5000 Gt obtained by S06 (it takes into account the link between the total meridional flux and mass flux in the western portion of the Pacific) gives an estimate of the typical size of the signal arriving in the tropical Pacific. This signal is substantial, as a positive (negative) mass change corresponds to thermocline elevation (depression) in the tropics about 50 m over an area of 10 degree by 100 degrees in the period of 3 months. This signal propagates across the equatorial Pacific by means of short time scales of Kelvin waves and influences the tropical SST via the deepening (shallowing) of thermocline depth. Thus these wave interactions can define the interannual SST fluctuations in the tropics via the variability of dynamics in the Southern Ocean, i.e., as it was shown by S06, these interannual SST fluctuations are associated with the interannual variability of atmospheric conditions over the ACC in the preceding 4-6 months, which in turn, is initiated by the variability in the tropics in the preceding couple months. S06 proposed a plausible explanation of coupled interaction of the tropics and high latitudes and their influence on ENSO events which includes the following processes. Upper ocean warming (cooling) in the tropics (that, e.g., could be associated with a seasonal cycle) leads to an enhanced (decreased) heating in the upper troposphere over the tropical ascending region in the Pacific. It means warmer (colder) air is transferred by the Hadley cell from the tropics to the descending regions in the subtropics that slows down (speeds up) here the atmospheric downwelling which then weakens (strengthens) wind over the Southern Ocean. As it was mentioned earlier (due to the anticorrelation between mass variations in the Pacific and the strength of wind over the ACC), the weak (strong) wind over the Southern Ocean is associated with equatorward (poleward) mass flux in the Southern Ocean to the north in the vicinity of 47°S that leads to the amplification of a warm (cold) ENSO signal via propagation of pressure/density anomalies from the Southern Ocean to the tropics by means of fast wave processes.

The interaction between the tropics and the Southern Ocean depends on the stochastic processes of ocean-atmospheric interactions in these regions. A substantial role in these stochastic processes, as it was shown by S06, can be due to the mass flux variability in the Southern Ocean associated with the changes in atmospheric forcings over the ACC, and would the interaction between the tropics and high latitudes lead to the ENSO event or usual seasonal variability, depends on the processes in the Southern Ocean.

Burgers et al. 2005, henceforth BJO05, presented the simplest form of the ENSO recharge oscillator model which is based on two equations: one for the eastern Pacific sea surface temperature anomaly $T_E$ and the other for the mean equatorial Pacific thermocline depth anomaly $h$ with the damping on $T_E$ much stronger than the damping on $h$. The interaction between $T_E$ and $h$ is characterized by the time delay between the east and the west of the Pacific that is due to both finite Kelvin wave speed and SST dynamics. In this paper authors have taken into account a parameterization of the fast wave process by which the thermocline tilt adjusts to the wind stress into the recharge oscillator. The parameters of the recharge oscillator model were obtained from two different methods, but both are based on a standard fit, that minimizes the rms error of model forecasts for the model variables and the observed values. This fit gives an oscillation period T and decay time $\gamma^{-1}$ of about 3 and 2 years, respectively, since the observed periods of ENSO lie in the range from 2 to 4 years.

In reality, as it was described earlier, the density anomalies defining the ocean dynamics in the vicinity of the equator propagate very fast meaning that an equatorial Kelvin wave would take about 2 months to cross the whole Pacific and therefore, any density/pressure disturbances appeared in western tropics will be revealed in eastern ones very quickly by means of propagation of internal (baroclinic) Kelvin waves (see, e.g., Blaker et al. 2006). Hence we can directly investigate the tropical variability at short periods by using the shorter period and decay time parameters than in BJO05 assuming that the propagation of short time scales of Kelvin waves and associated SST reaction lead to rise to interannual fluctuations that are only dependent on external forcing describing the variability of dynamics in the Southern Ocean and in the tropics.

## 2. Method and Results

The recharge oscillator of BJO05 is based on two equations for the $T_E$ and the $h$. A third equation describing the variability of bottom water thickness anomaly in the tropical western Pacific, $z$, is added here to parameterize the variability of the thermocline depth $h$ due to the fluctuations of meridional fluxes in the Southern Ocean. For this case, the equations of BJO05 system can be rewritten as:

$$\partial_t \boldsymbol{T_E} = -2\boldsymbol{\gamma}\, \boldsymbol{T_E} + \boldsymbol{\omega_o}\, \boldsymbol{h}, \qquad (1)$$

$$\partial_t \boldsymbol{h} = -\boldsymbol{\omega_o}\, \boldsymbol{T_E} - 2\boldsymbol{\gamma_B} h + \omega_B\, z, \qquad (2)$$

$$\partial_t z = -2\gamma_B z + F_{ex}, \qquad (3)$$

where $T = 2\pi\omega^{-1}$ and $T_B = 2\pi\Omega^{-1}$ are the periods, and $\gamma^{-1}$ and $\gamma_B^{-1}$ are the decay times describing the processes in the tropics and the middle latitudes respectively; $\omega^2 = \omega_o^2 - \gamma^2$; $\Omega^2 = \omega_B^2 - \gamma_B^2$; and $F_{ex}$ denotes the external forcing. Bold font in (1) - (2) denotes the terms of the original BJO5 model. The terms with subscript "B" describe the processes of interaction between the Southern Ocean and the tropics, that are due to fast-moving barotropic Rossby wave processes from the Southern Ocean to low latitudes (see IZD04) and they are added in equations (1)-(3) similar processes for the tropics that will be described below.

The external forcing added in the equation (3) is defined as being proportionate to the scaled monthly averaged mass variability of *M(t)* in the Pacific Ocean due to meridional transport fluctuations through the latitude of $40^oS$ (i.e., | *M(t)*|≤1) from 1985 to the end of 2004 (S06) with the coefficient $C_o$ ($F_{ex}= C_o$ *M(t)*). In the model this forcing describes the short period mass variability in the Pacific due to meridional transport fluctuations through the latitude of $40^oS$ which lead to the appearance of a cold (warm) temperature anomalies in the Southern Ocean (described by $F_{ex}$ in (3)), that are transferred to low latitudes by the wave mechanism by IZD04 (having the time scale of $T_B$). This increases (decreases) the bottom water thickness in the western Pacific, *z*, that leads to the elevation (depression) of the thermocline in the west equatorial Pacific. This mass surplus (lack) near the equator begins to disperse eastward as a so-called downwelling (upwelling) Kelvin wave resulting in deepening (shallowing) of the mean thermocline depth via the last term $\omega_B$ *z* in (2) (it is similar to the term $\omega_o$ **h** in (1) which describes the dependence of $T_E$ variability on the mean thermocline depth (with the time scale of T)). The terms with $\gamma_B$ define damping processes.

Thus, the terms of equations (1)-(3) with subscript "B" properly describe processes in the model: the high (low) value of $F_{ex}$ (i.e., *M(t)*) leads to an increase (decrease) of variable *z* that increases (decreases) *h*, and $T_E$ increases (decreases) too. The periods and decay times in the experiments were varied across a broad range: from 1 till 10 for the period and from 1 till 36 months for the decay time.

The SST anomalies averaged over the region between latitudes 5°S and 5°N, and between longitudes 160°E and 150°W of the Pacific (NINO4-index from http://climexp.knmi.nl) have been used for a comparison with the model $T_E$.

The statistical significance of all correlation coefficients presented in the paper is statistically significant at the 99% level that was determined via an effective sample size following Bretherton et al. (1999).

The scaled values for the $M(t)$ and the NINO4-index for the model's period are presented in Figure 2a. The correlation coefficient between the monthly averaged values of $M(t)$ and NINO4 is 0.27. The positions of major peaks for these curves are consistent, though the $M(t)$ curve displays more high frequency fluctuations than NINO4.

The equation system (1)-(3) was solved numerically from 1985 onwards with the initial conditions: $T_E|_{t=1985}=0$, $h|_{t=1985}=0$ (henceforth called experiment E1). With model parameters T=2 months, $\gamma^{-1}$ = 7 months, and $T_B$=5 months, $\gamma_B^{-1}$ = 10 months the oscillations with periods corresponding to ENSO were excited in the system. The value of parameter $C_o$=1.3 was chosen in the experiment that corresponds to the maximum amplitude of variability of $T_E \sim$ 1°C. The upper solid and the middle lines on Figure 2a correspond to the scaled model SST $T_E$ and thermocline depth $h$ obtained in this experiment. The oscillation of $h$ leads that of $T_E$ by about 2 months and the majority of its variability is due to the $M(t)$: the correlation of $M(t)$ with $h$ is 0.84. The correlation between model $T_E$ and NINO4 is 0.68. The percentage of variance explained was calculated to be about 43%. Figure 2b shows winter's $T_E$ and the NINO4-index (averaged during the three months from December to February when the warm

or cold ENSO events usually achieve the maximum phase of their development) and, for comparison, the preceding scaled summer's mass variability of *M(t)* in the Pacific Ocean due to meridional transport fluctuations through the latitude of 40°S. The latter is from the model's time series (S06), averaged from July to September. The correlation coefficient between winter's $T_E$ and the NINO4-index is 0.72 and the percentage of variance explained is about 48%.

Experiments in which the values of the model parameters were varied show that the model results have little dependence on the choice of the period $T_B$ or the damping coefficient $\gamma$, but they are sensitive to the decrease of $\gamma_B^{-1}$ and the increase of T. The correlation between $T_E$ and NINO4 decreases with increasing of T (decreasing $\gamma_B^{-1}$) and drops to about 0.5 when T ($\gamma_B^{-1}$) equals to 6 (5) months. An increase in the parameter $\gamma^{-1}$ leads to noisier behaviour of the variable *h*, but these changes are not significant for $T_E$.

It is seen clearly from Figure 2b that the warm ENSO events of 1986-87, 1991-92, 2003 and partly for 1997, along with the cold ENSO events of 1988-89 and 1998-2000 can be reproduced by this simple model. The warm and cold ENSO events occurred when the maximums and minimums of the ocean model's summer meridional flow from the Southern Ocean were observed, so the joint effect of atmospheric conditions over the ACC and bottom topography in the Southern Ocean could be considered as the mechanism amplifying (or may be triggering, since there is no other external forcing in the model) ENSO events. However, the warm ENSO event in 1994-95 was omitted by the model that demonstrates the nature of ENSO events is more complicated than this simple model, due to the interaction between the ocean and atmosphere over a much broader area.

## 3. The effect of westerly wind variability in the model

It has been established that the onset of ENSO depends on equatorial wind anomalies in the western Pacific during the preceding spring and summer, though these wind anomalies can trigger the ENSO when the oceanic conditions in the tropical Pacific are favourable to the development of the ENSO (see e.g. Lengaigne et al., 2004). ). It can also be seen from Figure 2 that the short-period meridional mass variability in the Southern Ocean can be considered as a "favourable" condition to set up the ENSO. Figure 3 shows the correlation between the wind stress zonally averaged over the Pacific from the ECMWF reanalysis, $<\tau_x>$, (June-September average) and winter's NINO4-index (averaged December-February). There are high correlations between ENSO and winds in the tropics, in the Trade Wind region and over the ACC. The interpretation of these high correlations is that weak winds in southern hemisphere set up warm ENSO events (for clarity the dashed line on Figure 3 represents the scaled profile of time averaged $<\tau_x>$). As mentioned above, there is an anticorrelation between the strength of the wind over the ACC and the variability of meridional mass fluxes in the Pacific which in turn, is significantly correlated with winter's NINO4-index in the latitude band from 45° to 35° S (dotted line in Figure 3). From the value of correlation coefficient between the mean summer's $M(t)$ and winter's NINO4 (~0.8), it can be calculated that the mean summer's $M(t)$ describes about half of winter's NINO4-index variance. Thus the variability of wind over the ACC in the ENSO model is taken into account via $M(t)$. To account for the effect of westerly wind variability in the tropics, the SOI-index will be used in the following experiments.

The SOI-index is defined as the normalized atmospheric pressure difference between Tahiti and Darwin, i.e. the higher SOI-index, the stronger tropical wind. There are several slight variations in the SOI values calculated at various centres. In following experiments the series from 1950 onwards calculated by the method of Ropelewski and Jones (1987) (obtained from the website http://climexp.knmi.nl) and data by Trenberth (http://www.cgd.ucar.edu/cas/catalog/climind, where the standardizing is done using the approach outlined by Trenberth (1984) to maximize the signal-to-noise ratio) were used. However, model results slightly depend on the choice of these data.

The additional external tropical force $F_T$ which parameterizes the effect of westerly wind variability was added to the right side of equation (2) of ENSO system model (1)-(3) which can be written in form:

$$F_T(t, \Delta\tau_T) = -C_T/\Delta\tau_T \int_{t-\Delta\tau_T}^{t} SOI(t)\, dt \equiv C_T/\Delta\tau_T \int_{t-\Delta\tau_T}^{t} SOI^-(t)\, dt, \qquad (5)$$

where $C_T$ is a scale factor; SOI is the scaled value, i.e. $|SOI|\leq 1$. The SOI contains both the long time scale component of the ENSO signal and noise components. To minimize the influence of noise on model results, the SOI averaged on the preceding interval $\Delta\tau_T$ is used. Besides, the $\Delta\tau_T$ should be chosen in such a way that it is greater than the time delay due to the propagation of the signal in the tropics as the thermocline tilt adjusts to the variability of wind stress by means of wave processes. The SOI-index is defined as having the opposite sign to $F_T$ from convention since it corresponds to setting up a warm ENSO event when the wind in the tropics is weakened. This negative SOI is denoted $SOI^-$.
Hence equation (2) becomes:

$$\partial_t h = -\omega_o T_E - 2\gamma_B h + \omega_B z + F_T, \qquad (6)$$

This modified equation system (1), (3) and (5)-(6) was solved numerically (henceforth E2) in a similar way to experiment E1. The value of parameter $C_T$ =5.1 was chosen in this experiment, so that the maximum amplitude of variability in $T_E$ corresponds to 1.6°C and so the contribution of both external forcings to this variability would be comparable (that follows from the correlation coefficient between the *M(t)* and NINO4-index). The correlation between SOI⁻ and NINO4 for the period from 1985 to 2005 is 0.38 (though for 1950-2005 this coefficient is 0.57), increasing up to 0.62 at 4 months lag, where the SOI⁻ leads the NINO4-index (0.70 for 1950-2005). On this basis the parameter $\Delta\tau_T$=4 months was chosen and after that the correlation between the forcing $F_T(t, \Delta\tau_T=4)$ and NINO4 increased.

The solution of the modified ENSO system is presented in Figure 4. All warm and cold ENSO events (including the warm ENSO event in 1994-95 omitted before) are now reproduced by this simple model. The analysis of the SOI-index and *M(t)* for 1994 shows that the previous failure to reproduce the ENSO of 1994-95 in experiment E1 was due to the presence of weak winds in southern hemisphere having low variability on time almost continuously during whole of 1994, that minimizes the joint effect of the variability of atmospheric conditions over the ACC on ocean dynamics in the Southern Ocean. However, the long-term weakness of westerly winds in the tropics leads to the onset of ENSO, which is now taken into account by the parameterization of the external tropical forcing. The correlation coefficient between the model $T_E$ and NINO4 is 0.83. The percentage of NINO4 variance explained by $T_E$ is more than 65%. The correlation coefficient between winter's $T_E$ and NINO4-index is 0.87, and the percentage of variance explained in this case is about 76%. Note that for the case of $\Delta\tau_T$=0 the correlation between the model $T_E$ and NINO4-index is 0.72 and the percentage of NINO4 variance explained by $T_E$ is 46%, i.e., the model results are similar to the results of experiment E1, when only forcing due to the effect of the Southern Ocean was taken into account. The choice of a larger scale factor $C_T$ for the model forcing due to the SOI, to obtain results comparable with the Southern Ocean contribution to variability in the eastern Pacific SST anomaly, demonstrates that the variability of ocean dynamics in the Southern Ocean makes a major contribution to the variability of tropical SST.

Experiments in which the values of the model parameters (T and γ) were varied demonstrated a similar dependence to experiment E1. The variation of $\Delta\tau_T$ (±2months) from $\Delta\tau_T$ =4 months has no significant effect on the correlation between $T_E$ and NINO4-index. Thus, this simple ENSO model is able to reproduce ENSO events very well.

## 4. The simplified forecast ENSO model

The modified system of the ENSO model (1), (3) and (5)-(6) can be reduced for the forecast of ENSO events (similar to the model of BJO05) by the following representation of the expression for external forcings in the Southern Ocean:

$$\partial_t h = -\omega_o T_E + F_{ex} + F_T, \qquad (7)$$

where

$$F_{ex}(t,\Delta\tau) = C_o /\Delta\tau \int_{t-\Delta\tau}^{t} M(t)\, dt. \qquad (8)$$

Here $C_o$ is a scale factor and $\Delta\tau$ is the time delay due to the propagation of the signal from the Southern Ocean to the equator (IZD04). Thus, the effects of variability due to both *M(t)* and the westerly wind variability in the tropics averaged for the previous $\Delta\tau$ and $\Delta\tau_T$ months will be used in the following numerical experiments.

Relying on IZD04's estimate of 4-6 months as the time needed for anomalies from the Southern Ocean to reach the low latitudes of the Pacific, the parameter $\Delta\tau$ in experiments was varied from 1 to 6 months. As in experiment E2, the parameter $\Delta\tau_T$=4 months was used. The values of the scaling factors $C_o$ =6.3 and $C_T$ =12.6 were also adopted in this experiment so that the maximum amplitude of the total variability of $T_E$ is 1.6°C and the contribution of both external forcings is comparable (it is about 1°C for both $F_T$ and $F_{ex}$). The parameters T and $\gamma$ in these experiments varied as in experiment E1

The equation system (1), (5) and (7)-(8) was solved numerically from 1951 onwards with the initial conditions $T_E|_{t=1951+\Delta\tau}$=0, $h|_{t=1951+\Delta\tau}$=0 (henceforth experiment E3). A time shift in the initial conditions is determined by the lag between the external forcing in the Southern Ocean (see Fig. 1) and in the tropics (Fig. 3), and the onset of an ENSO event. The force $F_{ex}$ describing the effect of dynamic variability in the Southern Ocean, appeared in the system after 1985 that due to available model data for $M(t)$. With model parameters T=2 months, $\gamma^{-1}$ = 7 months, and $\Delta\tau_T=\Delta\tau$=4 months the oscillations with the periods corresponding to ENSO were established in the system and reproduced all warm and cold ENSO events.

The scaled values for SOI⁻ (with a time lag of 4 months) and NINO4-index for the period from 1951 to 2005 are shown in Figure 5a. The behaviour of these curves are very similar (correlation coefficient is 0.70) though the original SOI⁻ curve contains more noise than NINO4, that is natural for atmospheric pressure variability in comparison with SST. The upper (solid) and middle lines on Figure 5a correspond to the model scaled values of SST, $T_E$, and thermocline depth $h$ obtained in experiment E3 for the period from 1951 to 2005. The variability in $T_E$ and $h$ is observed to increase after 1985 when $F_{ex}$ is included. Figure 5b shows the curves of $T_E$ and $h$ from 1985 in more detail. It can be seen that the oscillation of $h$ leads that of $T_E$ by about 2 months, similar to experiment E1.

The correlation coefficients for SOI⁻ and $M(t)$ with model $T_E$ at 4-months lag, for the period of 1985-2005, are about 0.65 (SOI⁻ and $M(t)$ lead $T_E$). In comparison, the correlation coefficient obtained for this period between the model $T_E$ and NINO4 is 0.82 (0.78 for 1951-2005). The percentage of variance explained is about 60% which is slightly less than in experiment E2. The correlation between the winter $T_E$ and the NINO4-index for 1985-2005 is 0.92 and the percentage of variance explained is about 84% that is slightly better than the results of experiment E2. Thus, this simple ENSO model is able to forecast ENSO events for 4 months in advance by using the model $M(t)$ and SOI-index averaged for the previous 4 months.

Experiments in which the values of the model parameters (T and γ) were varied demonstrated a similar dependence to experiment E1. Variation of Δτ from Δτ=4 months slightly decreases the correlation coefficient between $T_E$ and NINO4-index, decreasing by about 0.2 and 0.1 for Δτ=2 and Δτ=6, respectively. Hence, the Δτ=4 months is the optimum choice.

## 5. Summary

A modified version of the simple model of BJO05 which is a classical damped oscillator, with eastern Pacific SST and the mean equatorial Pacific thermocline depth representing momentum and position respectively, was used to model ENSO events. The main difference between the original BJO05 and the modified model is the presence of external forcings and the use of shorter period and decay time parameters meaning that the propagation of short time scales of Kelvin waves and associated SST reaction lead to rise to interannual fluctuations that are completely dependent on external forcings describing the variability of dynamics in the Southern Ocean and in the tropics. The external forcings in the model are parameterized by the short period mass variability in the Pacific sector of the Southern Ocean due to meridional transport fluctuations through the latitude of $40^{o}$S, and the SOI$^{-}$ index (defined with the opposite sign to convention). The first forcing describes the variability due to the joint effect of the atmospheric variability over the ACC, bottom topography and coastlines in the Southern Ocean; the second forcing describes westerly wind variability in the tropics. Both forcings are results of coupled interactions between the tropics and high latitudes. Under such conditions oscillations arise in the ENSO system as a result of propagation of signals due to both initial signals appeared in the Southern Ocean and the tropical westerly wind anomaly, that propagate then across the equatorial Pacific by means of fast wave processes.

The external forcings are the main factor in the establishment of the oscillation pattern in the ENSO forecast. To obtain results comparable with the Southern Ocean contribution to variability in the eastern Pacific SST anomaly, a larger scale factor for the model forcing due to the weakness of westerly winds in the tropics was chosen. It demonstrates that the variability of ocean dynamics in the Southern Ocean makes a major contribution to the variability of tropical SST though it is initiated by the variability in the tropics in the preceding couple months. However, the westerly wind variability in the tropics becomes more important when weak westerly winds established in the tropics during very long periods (about year) leading to the onset of ENSO, while weak winds in southern hemisphere having low variability on time minimize the joint effect of the variability of atmospheric conditions on ocean dynamics in the Southern Ocean.

It was shown that this simple ENSO model is able to forecast the ENSO events for 4 months in advance by using the short period model mass variability in the Pacific sector of the Southern Ocean due to transport fluctuations through its open boundary, along with the SOI-index, each averaged over the previous 4 months. A model skill of 0.92 for a four-month lead forecast of the December-February ENSO is comparable with the correlation between August-September NINO4-index and the subsequent December-February NINO4-index and it may be seemed not so impressive. The most important point of these model results is the establishment of two major and comparable feedbacks (in a system with so many feedbacks and connections) responsible for the onset of ENSO, accounting for about 84% of the percentage of variance explained:

- short-period meridional mass fluctuations in the Pacific sector of the Southern Ocean due to the joint effect of the atmospheric variability over the ACC, bottom topography and coastlines;
- the variability of westerly winds in the tropics.

**Acknowledgments**. This work was funded by the Natural Environment Research Council. Thanks to Simon Holgate for commenting on this manuscript.

# References

1. Alvarez-Garcia F., W.C. Narvaez, and M.J. Ortiz Bevia (2006) An assessment of differences in ENSO Mechanisms in a Coupled GCM Simulation, J. Climate, 19, 69-87.
2. Blaker, A.T., B. Sinha, V.O. Ivchenko, N.C. Wells, and V.B. Zalesny (2006), Identifying the roles of the ocean and atmosphere in creating a rapid equatorial response to a Southern Ocean anomaly, Geophysical Research Letters, v.33, L06720, doi:10.1029/2005GL025474.
3. Bretherton, C. S., M. Widmann, V.P. Dymnikov, J. M. Wallace, and I. Bladě (1999), The effective number of spatial degrees of freedom of a time-varying field, J. Clim., 12(7), 1990-2009.
4. Burgers G., F.-F. Jin, and G.J. van Oldenborgh (2005), The simplest ENSO recharge oscillator, Geophysical Research Letters, v.32, L13706, doi:10.1029/2005GL022951.
5. Ivchenko V.O., V. B. Zalesny, and M.R. Drinkwater (2004), Can the equatorial ocean quickly respond to Antarctic sea ice/salinity anomalies?, Geophysical Research Letters, v.31, L15310, doi:10.1029/2004GL020472.
6. Jin, F.-F., (1996), Tropical ocean-atmosphere interaction, the Pacific Cold Tongue, and the El Niño/Southern Oscillation, Science, 274, 76-78.
7. Kessler, W.S. (2002), In ENSO a cycle or series of events? Geophys. Res. Lett., 29 (23), 2135, doi:10.1029/2002GL015924.
8. Lengaigne M., E. Guilyardi, J.-P. Boulanger, C. Menkes, P. Delecluse, P. Inness, J.Cole, J. Slingo (2004), Triggering of El Niño by westerly wind events in a coupled general circulation model. Climate Dynamics, 23, 601-620, doi:10.1007/s00382-004-0457-2.
9. Richardson G., M. R. Wadley, and K. Heywood (2005), Short-term climate response to a freshwater pulse in the Southern Ocean, Geophysical Research Letters, v.32, L03702, doi:10.1029/2004GL021586.
10. Ropelewski, C.F. and Jones, P.D. (1987) An extension of the Tahiti-Darwin Southern Oscillation Index. Monthly Weather Review, 115, 2161-2165.
11. Stepanov V.N., and C.W. Hughes (2004), The parameterization of ocean self-attraction and loading in numerical models of the ocean circulation, J. Geophys. Res., 109, C0037, doi:10.1029/2003JC002034.
12. Stepanov V.N., and C.W. Hughes (2006) Propagation of signals in basin-scale bottom pressure from a barotropic model, J. Geophys. Res., 111, C12002, doi:10.1029/2005JC003450.
13. Suarez, M., and P.S. Schopf (1988), A delayed action oscillator for ENSO, J. Atmos. Sci., 45, 3283-3287.
14. Trenberth (1984), Signal versus Noise in the Southern Oscillation, Monthly Weather Review 112:326-332

**List of Figure Captions**

**Fig. 1** - The values of transport through Drake Passage in Sv (thin solid line) and daily mass variability $M(t)$ in the Pacific Ocean due to meridional transport fluctuations through the latitude of 40°S in Gt (Gigatonns) (thick solid line) averaged for July-September. Symbols EL and LA denote warm and cold ENSO events, respectively. Dashed line corresponds to scaled winter's NINO4-index.

**Fig. 2 a** – Mass variability $M(t)$ due to meridional transport fluctuations through 40°S in the Pacific Ocean (the lowest line), scaled by a maximum of its value, the thermocline depth anomaly, $h$ (middle line), the SST anomaly, $T_E$ (upper solid line) and the NINO4-index (upper dashed line) for the period from 1985 to 2005; the value of NINO4-index is scaled by a factor of 1.6; **b**- the winter $T_E$ (solid line) and NINO4-index (averaged from December to February) (dashed line) and the preceding summer's $M(t)$, averaged from July to September (dashed-dotted line).

**Fig. 3** The correlation coefficient: dotted line for the meridional summer's mass fluxes averaged from July until September and solid line for the zonal wind stress over the Pacific averaged from June to September, each with winter's NINO4-index (averaged during three months from December until February). The dashed line represents the scaled profile of time averaged zonal wind stress. The positive sign corresponds to the eastern direction.

**Fig. 4** As Figure 2 but for experiment E2.

**Fig. 5 a** - the scaled values for SOI⁻ index with a 4 months time lag (the lowest line), the thermocline depth anomaly, $h$ (the middle line), the SST anomaly, $T_E$ (upper solid line) and the NINO4-index (the upper dashed line) for the period from 1951 to 2005; the value of NINO4-index and $T_E$ are scaled by a factor of 1.6; **b**- the scaled winter's $T_E$ and NINO4-index with preceding mean summer's model $M(t)$ and SOI⁻ index (top); $h$, $T_E$ and the NINO4-index for the period from 1985 to 2005 (bottom).

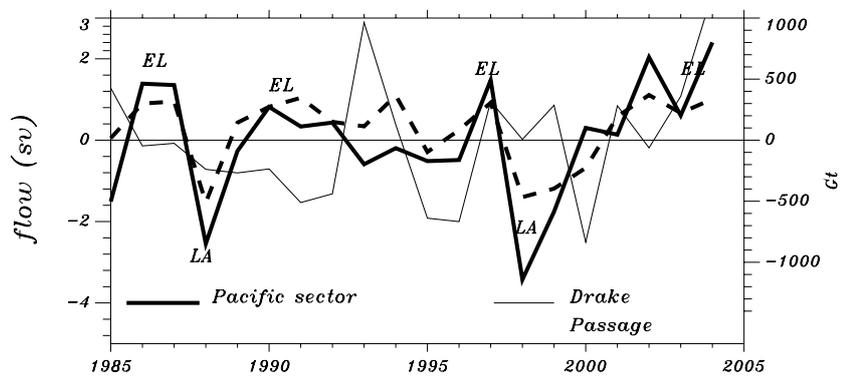

Figure 1

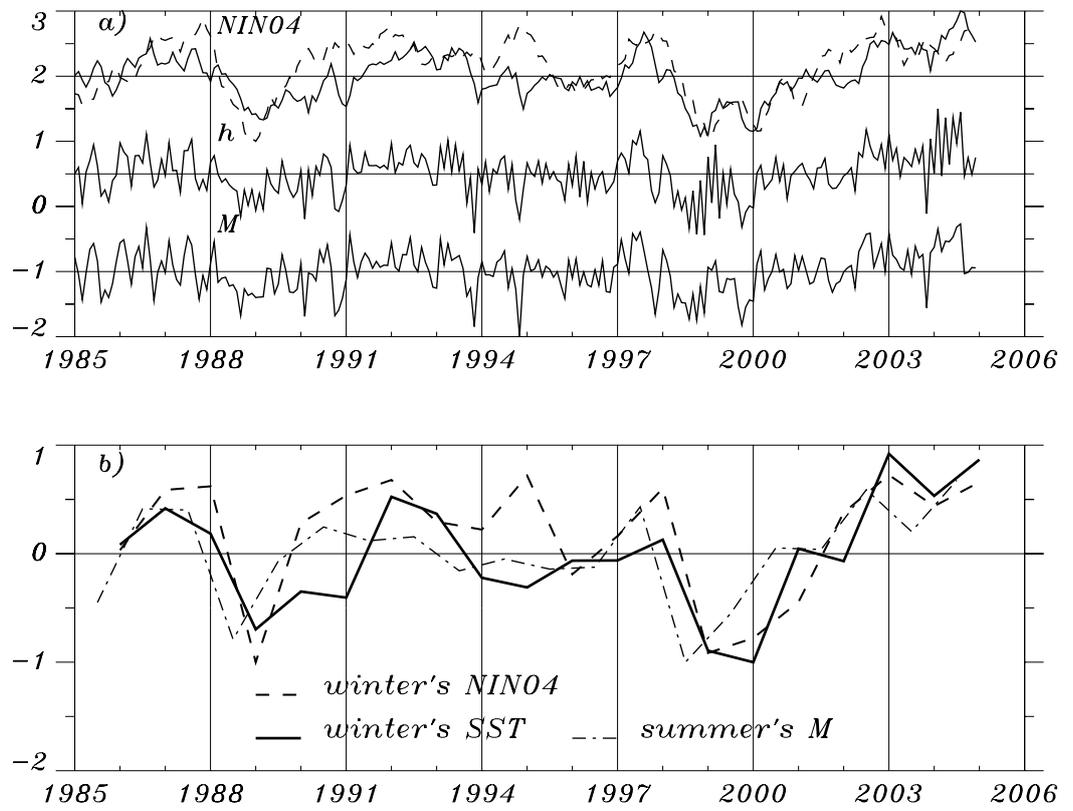

Figure 2

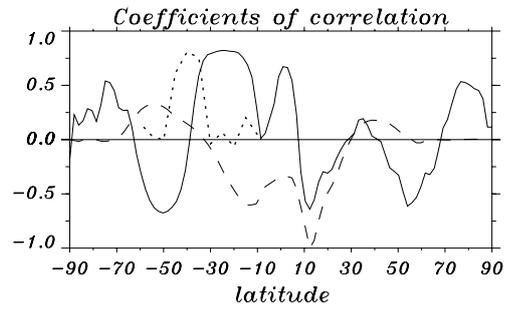

Figure 3

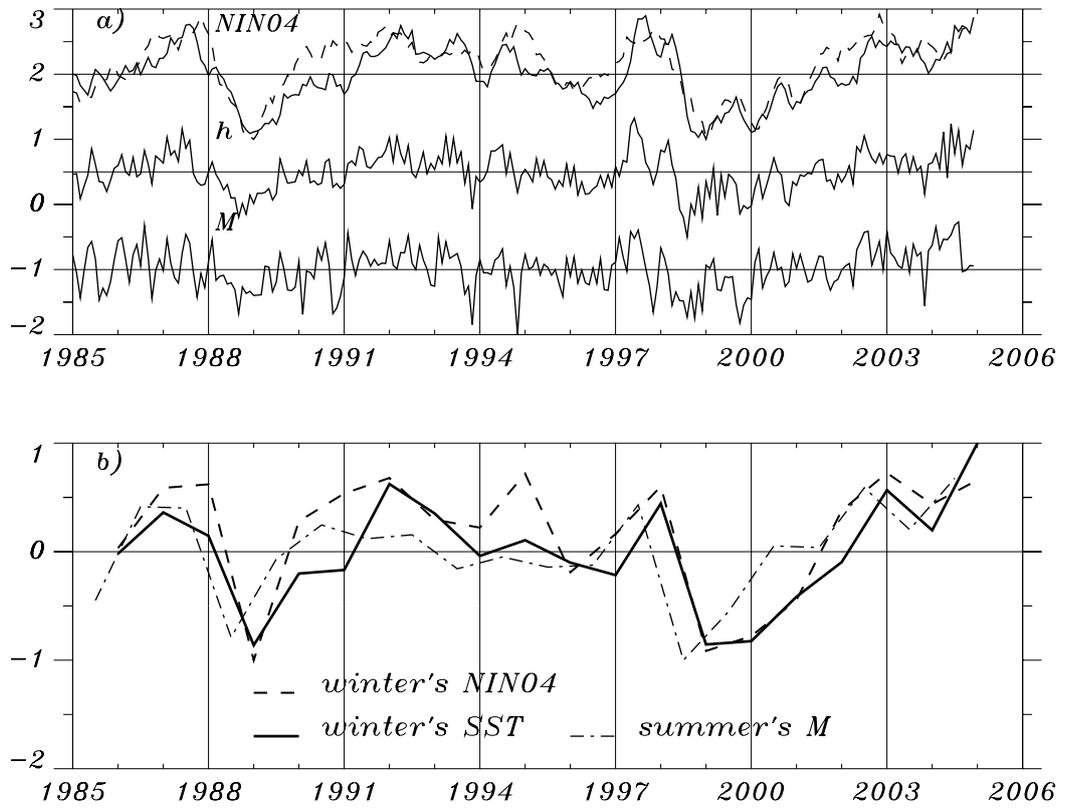

Figure 4

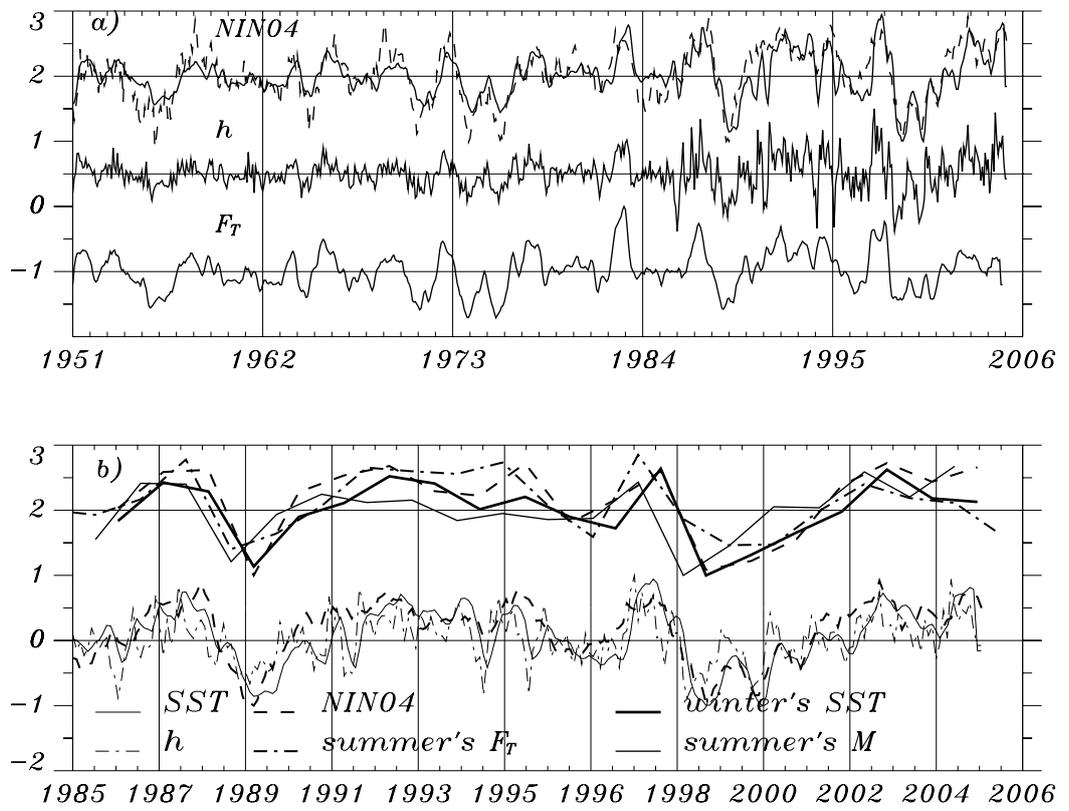

Figure 5